\documentclass[aps,prb,floats,
twocolumn,nobibnotes]{revtex4}
\usepackage[dvips]{graphicx}
\usepackage{amsmath}
\usepackage{bm}

\begin{document} 
 
\title{
Exact mapping 
of periodic Anderson model to Kondo lattice model
}
\date{\today}  

\author{P.~Sinjukow}
\email{peter.sinjukow@physik.hu-berlin.de}
\author{W.~Nolting}
\affiliation{Lehrstuhl Festk{\"o}rpertheorie, Institut f{\"u}r Physik,
  Humboldt-Universit{\"a}t zu Berlin, Invalidenstr.\ 110, 10115 Berlin}

\begin{abstract}
It is shown that the Kondo lattice model, for any finite coupling
constant $J$, can be obtained exactly from the periodic Anderson 
model in an appropriate limit. The mapping allows a direct proof of
the ``large'' Fermi volume for a nonmagnetic Fermi-liquid state
of the Kondo lattice model. 
\end{abstract}

\maketitle

The periodic Anderson model (PAM) and the Kondo lattice model (KLM) 
belong to the most intensively discussed many-body 
models in solid state physics. 
Both are believed to at least qualitatively describe important aspects
of the extremely rich physics of so-called 
heavy-fermion systems \cite{Hew,Faz99,TSU97}.
The properties of those systems are based on an interplay between 
rather localized $f$ electrons and
itinerant $s$, $p$ or $d$ conduction electrons.
In the (non-degenerate) periodic Anderson model this is accounted for 
in a minimal way, considering non-degenerate $f$ orbitals with 
intra-orbital Coulomb interactions and a non-degenerate conduction 
band with which the $f$ orbitals are hybridized. The effective physics 
is often described in terms of the Kondo lattice model, where 
the $f$ electrons are modelled as localized quantum-mechanical spins 
with an antiferromagnetic spin exchange with the conduction electrons
via a coupling constant $J$.  
The KLM represents an effective model of the PAM in the so-called 
Kondo regime, which is relevant to heavy-fermion systems.
In the Kondo regime the correspondence between the models 
is an approximate, perturbational one.
It is known to become exact
in the so-called Kondo limit of the PAM, which corresponds to the
weak-coupling limit ($J\to 0$) 
of the KLM.

This relation between the KLM and the PAM was established 
a long time ago with the help of the Schrieffer-Wolff
transformation \cite{ScW66,LaC79,PrL81},
which was originally devised for the corresponding impurity models
(single-impurity Anderson model \cite{And61} and Kondo impurity model
\cite{Kon64}) to show that those are related in a similar way.
In a more recent paper Matsumoto and Ohkawa claimed for the 
single-impurity Anderson model and the Kondo impurity model 
an equivalence in a special ``$s$-$d$ limit'', which differs from the 
Kondo limit \cite{MaO95}. Based on that, they inferred
the same equivalence to hold between the PAM and the KLM in the
case of infinite spatial dimensions. 
In this Letter we show that Matsumoto's and Ohkawa's $s$-$d$ limit,
in the following also called ``extended Kondo limit'', can be used for
a \textit{direct} and \textit{rigorous} mapping of the periodic Anderson
model to the Kondo lattice model 
\textit{in any dimensions}.
Thus a fundamental relation between both models is established.
In contrast with the (conventional)
Kondo limit, the equivalence in the extended Kondo limit rigorously
holds for any coupling $J$ of the KLM.

The general fact of an exact mapping of the PAM to the KLM
provides a general and rigorous answer to the long-standing
issue of the ``correct'' Fermi-surface sum rule for the
Kondo-lattice model. 
Luttinger's theorem \cite{LuW60}, 
which states that the volume enclosed by the Fermi
surface (``Fermi volume'') is ($a$) independent of the interaction
strength as long as no phase transition occurs and ($b$) 
otherwise only related to the number of electrons, 
cannot be directly applied to the KLM
since it is not a purely fermionic model. 
In particular it is a priori 
unclear, whether the localized spins can count as 
electrons in this context.
With the exact mapping the \textit{general} 
and \textit{rigorous} answer can be given: the respective 
sum rule of the PAM is rigourously mapped to the KLM.
We prove exemplarily that the Fermi volume of the 
KLM includes the number of localized spins (so-called ``large'' Fermi
volume) if the system is in a nonmagnetic Fermi-liquid state. 
We thus confirm in a much simpler way the result of a recent
topological proof by Oshikawa\cite{Osh00}.

It is clear that via the extended Kondo limit 
it will be possible in future to obtain further 
analytical and computational results
for the Kondo lattice model. 
Based on the rigorous mapping,
any result of the PAM translates into one for the KLM, and 
any analytical or computational method for the PAM
can be directly applied to the KLM.

This Letter is organized as follows: First, we review the
correspondence between the Kondo regime of the 
periodic Anderson model and the 
weak-coupling regime of the Kondo lattice model. 
Then, the proof of 
rigourous equivalence between the KLM and the PAM in the extended Kondo
limit is given. 
Finally, the direct proof of the large Fermi volume of the KLM which
follows from the exact mapping is explained.

The Hamiltonian of the periodic Anderson model (PAM) is given by
\begin{align}
H_{\mathrm{PAM}} =  \sum_{k\sigma} \epsilon_k n_{k\sigma} & +  \sum_{i\sigma} \epsilon_f
n_{i\sigma}^{f} + U \sum_i n_{i\uparrow}^f n_{i\downarrow}^f
\nonumber\\
& + \sum_{ki\sigma} 
( V_{k} e^{-i k R_i} \, c_{k\sigma}^\dagger
f_{i\sigma} + 
\mathrm{H.C.} 
) \;.
\label{eq:H_PAM}
\end{align}
$c_{k\sigma}^{(\dagger)}$ creates/annihilates a conduction 
electron ($s$ electron) with momentum $k$, spin $\sigma$ 
and one-particle energy $\epsilon_k$.
$f_{i\sigma}^{(\dagger)}$ is the creation/annihilation operator 
for an $f$ electron at site $R_i$ with energy $\epsilon_f$.
$U$ is the Coulomb repulsion between $f$ electrons 
at the same site (same $f$ orbital).
$s$-electron states are hybridized with $f$ orbitals  
via hybridization matrix elements
$V_k$.

The Hamiltonian of the Kondo lattice model (KLM) 
reads
\begin{align}
H_{\mathrm{KLM}}= \sum_{k\sigma} \epsilon_k n_{k\sigma} + \sum_{k k^\prime i}
J_{k^\prime k} e^{-i(k^\prime-k)R_i} \mathbf{S}_i \cdot
\mathbf{s}_{k^\prime k} \;.
\label{eq:H_KLM}
\end{align}
The first part describes the conduction band.
The second part stands for the interaction
between localized quantum-mechanical spins $\mathbf{S}_i$ of magnitude
$1/2$ and the spins
of the conduction electrons $\mathbf{s}_{k^\prime k}=
\frac{1}{2}\sum_{\sigma^\prime \sigma} c_{k^\prime \sigma^\prime}^{\dagger}
\bm{\tau}_{\sigma^\prime \sigma}  c_{k \sigma}$ 
(with $\bm{\tau}$ representing the Pauli matrices).  
$J_{k^\prime k}$ are the (antiferromagnetic) coupling constants 
($J_{k^\prime k}>0$).

The Kondo regime of the PAM is a regime favourable for the 
formation of local $f$ moments.
A necessary condition clearly is 
that the energy of a singly (doubly) occupied $f$
orbital lies below (above) the chemical potential: $\epsilon_f<0$,
$\epsilon_f+U>0$. The energy distance 
to the chemical potential should be large
compared with the hybridization 
so that fluctuations of $f$-orbital occupancy 
are small. This condition is usually 
formulated in terms of the width $\Gamma$ of the virtual level of the
single-impurity Anderson model \cite{TSU97,LaC79}:
\begin{align}
\frac{\Gamma}{|\epsilon_f|}, \;\frac{\Gamma}{\epsilon_f+U} \ll 1\;,
\end{align}
where $\Gamma=\pi \rho_0 V^2$. $\rho_0$ is the density of states of the
conduction band at the Fermi energy. $V$  is the average hybridization 
($V^2= \{ |V_k|^2 \}_\mathrm{av}$).
Via the Schrieffer-Wolff transformation, the Kondo regime of the PAM  
is approximately mapped to the weak-coupling (small-$J$) regime 
of the Kondo lattice model. 
Assuming a constant density of states,
the limit in which the mapping becomes exact (Kondo limit)
is given by
\begin{align}
\frac{V^2}{\epsilon_f}\,,\;\frac{V^2}{\epsilon_f+U} \longrightarrow 0\;.
\label{eq:Kondo_limit}
\end{align}
The Kondo limit can be understood as $V\to 0$ or $|\epsilon_f|,
\epsilon_f+U\to \infty$. Both corresponds to $J\to 0$ 
on the side of the Kondo lattice model \footnote{
As poined out 
by S.~K.~Kehrein and A.~Mielke (Ann.~Phys.~\textbf{252}, 1 (1996)), if 
$\epsilon_f$ or $\epsilon_f+U$ lie within the conduction
band (which may be the case if just taking the limit $V\to 0$),
the Schrieffer-Wolff transformation
is actually problematic 
because of energy denominators which become zero.
The problem does 
not occur in the extended Kondo limit.}.

The extended Kondo limit (EKL), 
which leads to an exact mapping of the
periodic Anderson model to the Kondo lattice model 
for arbitrary $J>0$,
is given by
\begin{align}
\epsilon_f&\equiv -\frac{U}{2} 
\nonumber\\
&U\to \infty \;\;,\;\;
V\to \infty 
\;\;\;\;\mbox{with}\;\;\;\;
\frac{V^2}{U}\to \mbox{const.}
\label{eq:EKL}
\end{align}
Note that $\epsilon_f \to-\infty$ as $U\to\infty$.
The proof of exact mapping in the EKL consists of two steps.
First, a finite Schrieffer-Wolff transformation 
is performed on the Hamiltonian of the PAM.
Second, the consequences of the EKL 
on the transformed Hamiltonian are checked to rigorously 
prove that the only terms which remain relevant are those of
the Kondo-lattice model.

The first three terms of $H_{\mathrm{PAM}}$ 
are denoted by $H_0$, the hybridization term by $H_V$. To eliminate 
all terms first-order in $V_k$, a unitary transformation $\bar{H} = e^S
H_{\mathrm{PAM}} e^{-S}$ is performed with the condition $[S,H_0]=-H_V$.
The required generator is 
\begin{align}
S= \sum_{k i \sigma} &\left( \frac{V_{k} e^{-i k R_i}}{\epsilon_k - \epsilon_f - U}
  n_{i-\sigma}^f c_{k\sigma}^\dagger f_{i\sigma} \right.
\nonumber\\
& \left. 
\hspace*{1em} + \frac{V_{k} e^{-i k R_i}}{\epsilon_k - \epsilon_f}
  (1-n_{i-\sigma}^f) c_{k\sigma}^\dagger f_{i\sigma} 
\right) 
- \mathrm{H.C.} 
\end{align}
The transformed Hamiltonian is given by
\begin{align}
\bar{H} = H_0 + 
H_2 
+ \frac{1}{3}[S,[S,H_V]]
 + \frac{1}{8}[S,[S,[S,H_V]]]+\ldots \;,
\label{eq:H_bar}
\end{align}
with 
\begin{align}
H_2\equiv \frac{1}{2}[S,H_V] 
\,=\, &H_{\rm ex} + H_{\rm dir} +
H_{\rm hop} + H_{\rm ch} \;,
\end{align}
where
\begin{align}
&H_{\rm ex} = \frac{1}{2} \sum_{kk^\prime  i} J_{k^\prime k}
e^{-i(k^\prime-k)R_i} 
\big(S_i^+ c_{k^\prime \downarrow}^\dagger c_{k\uparrow} + S_i^- c_{k^\prime
\uparrow}^\dagger c_{k\downarrow} + 
\nonumber\\
&\hspace*{11.5em} 
+S_i^z (c_{k^\prime \uparrow}^\dagger
c_{k\uparrow} - c_{k^\prime \downarrow}^\dagger c_{k\downarrow}) \big)
\label{eq:H_ex}
\\
&H_{\rm dir}= \!\! \sum_{kk^\prime i \sigma} \big( W_{k^\prime k} - \frac{1}{4}
J_{k^\prime k} ( n_{i\uparrow}^f + n_{i\downarrow}^f ) \big)
e^{-i(k^\prime-k)R_i}
c_{k^\prime\sigma}^\dagger
c_{k\sigma}\;
\\
&H_{\rm hop}  = - \sum_{kij\sigma} \big( W_{kk} 
 -  \frac{1}{4} J_{kk} 
(n_{i-\sigma}^f + n_{j-\sigma}^f) \big)
e^{-ik(R_i-R_j)} \ast
\nonumber\\
&\hspace*{18em} \ast f_{j\sigma}^{\dagger} f_{i\sigma}
\\
&H_{\rm ch} = - \frac{1}{2} \sum_{kk^\prime i \sigma} V_{k^\prime}
V_{k} e^{- i (k^\prime + k) R_i} \big( (\epsilon_{k^\prime} - \epsilon_f
-U)^{-1} -
\nonumber
\\
&\hspace*{4.5em} - (\epsilon_{k^\prime} -
\epsilon_f)^{-1} \big) 
c_{k^\prime-\sigma}^\dagger c_{k\sigma}^\dagger f_{i\sigma} f_{i-\sigma}
\hspace{0.1em}+ \mathrm{H.C.} 
\end{align}
with coupling constants 
\begin{align}
&J_{k^\prime k} = V_{k^\prime} V_{k}^\ast \{
-(\epsilon_k-\epsilon_f - U)^{-1} - (\epsilon_{k^\prime} -\epsilon_f - U)^{-1} 
\nonumber\\
&\hspace*{7em} + ((\epsilon_k-\epsilon_f)^{-1} + (\epsilon_{k^\prime} -\epsilon_f)^{-1} \}\;,
\\
&W_{k^\prime k}=\frac{1}{2} V_{k^\prime} V_{k}^\ast
\{(\epsilon_k-\epsilon_f)^{-1} + (\epsilon_{k^\prime} -
\epsilon_f)^{-1}\} \;.
\end{align}
The spin operators in (\ref{eq:H_ex}) are given by 
$\mathbf{S}_i=
\frac{1}{2}\sum_{\sigma^\prime \sigma} f_{i \sigma^\prime}^{\dagger}
\bm{\tau}_{\sigma^\prime \sigma}  f_{i \sigma}$.

Assuming a conduction band of finite width, 
the norm of the generator in the EKL has the asymptotics
\begin{align}
||S||\stackrel{\rm EKL}{\propto} \frac{V}{U}\;.
\label{eq:norm_generator}
\end{align}
With $||H_V|| \propto V$ and $V\stackrel{\textrm{EKL}}{\propto}
  \sqrt{U}$ 
it follows that all higher commutators in (\ref{eq:H_bar}),
starting at the order $V^3/U^2$, exactly vanish in the EKL,
\begin{align}
&[S,[S,H_V]]\;,\;\;[S,[S,[S,H_V]]] \;,\; \ldots \stackrel{\rm
  EKL}{\longrightarrow} 0 \;,
\label{eq:commut_EKL}
\end{align}
and it is sufficient to consider the EKL of the 
remaining Hamiltonian $\bar{H}^\prime \equiv H_0 + H_2$.

It is important to note that one cannot proceed
with 
the original argument 
given by Schrieffer and Wolff
for the Kondo regime of the single-impurity Anderson model \cite{ScW66}.
The reason is that
apart from $H_{\rm ch}$ also 
$H_{\rm hop}$ changes the number
of $f$ electrons at 
given sites.
$H_{\rm hop}$ connects the subspace of single $f$ occupancy with
the subspaces of zero and double occupancy. 
Therefore,
the Hilbert space cannot be separated at this stage.
To prove an effective fixing of $f$ occupation,
one needs 
to apply 
a different and more formal 
line of argumentation.

We denote the $s$ and $f$ electron parts of $H_0$ separately,
\begin{align}
H_0^s = \sum_{k\sigma} \epsilon_k n_{k\sigma} 
\;
,\;
H_0^U = 
\sum_{i\sigma} 
\epsilon_f 
n_{i\sigma}^{f} + U \sum_i n_{i\uparrow}^f n_{i\downarrow}^f \;.
\end{align}
In the EKL the different parts of $\bar{H}^\prime$
behave as:
\begin{align}
||H_0^s|| &\propto {\cal W} = \mathrm{const.}
\label{eq:EKL_Hprime1}
\\
||H_2|| &\propto \tilde{J} \equiv \frac{V^2}{U} = \mathrm{const.}
\\
||H_0^U|| &\propto U \stackrel{\mathrm{EKL}}{\longrightarrow}\infty \;.
\label{eq:EKL_Hprime3}
\end{align}
${\cal W}$ is the width of the conduction band.
Obviously, with respect to $\bar{H}^\prime$ 
the EKL is equivalent to just taking the limit $U\to\infty$ (and 
$\epsilon_f\equiv-\frac{U}{2}\to 
-\infty$).
$V$ needs not to be considered explicitely since it only
appears within the ratio ${V^2}/{U}\equiv \tilde{J}$, which is 
a constant in the EKL.

Let us consider $\bar{H}^\prime$ and its eigenstates 
as functions of the three parameters ${\cal W}$, $\tilde{J}$ and $U$.
It is clear that 
the eigenstates $\{|\Psi({\cal W},\tilde{J},U)\rangle\}$
actually only depend on the ratios ${\cal W}/U$ and $\tilde{J}/U$.
Therefore, in the EKL ($U\to\infty$) each eigenstate 
$\left|\Psi\right\rangle$ of $\bar{H}^\prime$ approaches an
eigenstate $\left|\Psi^0\right\rangle$ of $H_0^U$:
\begin{align}
\left|\Psi({\cal W}, \tilde{J}, U )\right\rangle \stackrel{U\to
  \infty}{\longrightarrow} 
\left|\Psi(0, 0, U^\prime
)\right\rangle \equiv  
\left|\Psi^0\right\rangle
\end{align}
with arbitrary $U^\prime$. Note that the states 
$\left\{\left|\Psi^0\right\rangle\right\}$
that are approached in the EKL are  
highly non-trivial
superpositions of trivial degenerate
eigenstates of $H_0^U$. Still, they can be grouped into 
two classes: first, states $\{\left|\Psi_1^0\right\rangle\}$ 
with a single $f$ electron at each site, and 
second, states $\{\left|\Psi_2^0\right\rangle\}$ 
with admixtures of zero and double $f$ occupation.
The energies of the $\left|\Psi_2^0\right\rangle$'s are higher
than the energies of the $\left|\Psi_1^0\right\rangle$'s by amounts
proportional to $U$. 
In the EKL ($U\to \infty$) the statistical weights of the
$\left|\Psi_2^0\right\rangle$'s obviously vanish. Moreover, 
the creation or annihilation of
$s$ electrons, which must be taken into account with regard 
to $s$-electron Green's functions of the KLM, 
do not connect the $\left|\Psi_1^0\right\rangle$'s
with the $\left|\Psi_2^0\right\rangle$'s. Hence, in the EKL
only the states $\{\left|\Psi_1^0\right\rangle\}$ are relevant.
The number of $f$
electrons at each site is effectively fixed to one: $n_{i\uparrow}^f +
n_{i\downarrow}^f=1$. 

Based on this, an
effective Hamiltonian $\bar{H}^{\prime\prime}$ can be formulated, which 
describes only the relevant states of $\bar{H}^\prime$ in the EKL.
Using $n_{i\uparrow}^f + n_{i\downarrow}^f=1$, several terms of
$\bar{H}^\prime$ can be 
neglected. 
Since $H_0^U$ is the only diverging term,
there 
cannot be 
any finite effective interactions 
that are 
omitted this way. 
$H_{\mathrm{ch}}$ can be neglected completely. $H_{\rm hop}$ 
reduces to the constant 
$-N\sum_{k} W_{kk}$, $N$ being the number of lattice sites.
The coupling constants simplify to 
\begin{align}
&W_{k^\prime k} \stackrel{\rm EKL}{=}  \frac{2 V_{k^\prime} V_{k}^\ast}{U}
\label{eq:Wkkprime_i}
\\
\mbox{and}\hspace*{2em} &J_{k^\prime k} \stackrel{\rm EKL}{=} \frac{8 V_{k^\prime} V_{k}^\ast}{U} \;.
\label{eq:Jkkprime_i}
\end{align}
Taking (\ref{eq:Wkkprime_i}) and (\ref{eq:Jkkprime_i}) into account,
$H_{\rm dir}$ exactly vanishes.
Neglecting the Hubbard term ($U \sum_i n_{i\uparrow}^f
n_{i\downarrow}^f$) as it describes double $f$ occupation,
the effective Hamiltonian in the EKL is 
finally given by 
\begin{align}
\bar{H}^{\prime\prime}= \sum_{k\sigma} \epsilon_k n_{k\sigma} + 
H_{\rm ex} + N \epsilon_f - N\sum_{k} W_{kk}
\;.
\label{eq:Hbar_end}
\end{align}
Apart from constants, $\bar{H}^{\prime\prime}$ corresponds to 
the Kondo lattice model $H_{\rm KLM}$. 
As there are no $f$-electron fluctuations, the
spin operators $\mathbf{S}_i$ in $H_{\rm ex}$ 
now describe localized quantum-mechanical spins of magnitude $1/2$.

As $V/U\stackrel{\rm EKL}{\longrightarrow}0$, for the generator 
one has $S \stackrel{\rm EKL}{\longrightarrow} 0$.
Therefore, the unitary Schrieffer-Wolff transformation 
reduces to an identical transformation.
Thus, in terms of relevant states
and disregarding unimportant constants, we have 
proven
\begin{align}
H_{\mathrm{PAM}} \stackrel{\rm EKL}{\longrightarrow} H_{\mathrm{KLM}} \;.
\label{eq:ident_trans}
\end{align}
The coupling constants of the KLM are
given by (\ref{eq:Jkkprime_i}).

Only very recently the long-standing issue of the large Fermi volume 
for a nonmagnetic Fermi-liquid state of the Kondo lattice model 
was solved by Oshikawa by means of a 
nonperturbative topological
proof of Luttinger's theorem \cite{Osh00}.
Oshikawa's result represents the first proof of the large Fermi volume
for arbitrary dimensions and coupling strengths after a 
number of special results had been achieved (variational
results \cite{ShF90}, proof for strong-coupling limit 
in one dimension \cite{UNT94}, proof for infinite dimensions 
\cite{MaO95}, general proof for one dimension \cite{YOA97}). 
The exact mapping of the periodic Anderson model 
to the Kondo lattice model in the extended Kondo limit
implies immediately another general proof, 
which is more direct than the one given by Oshikawa.

To be explicit,
for the PAM the Luttinger theorem \cite{LuW60} states that
the Fermi volume of a nonmagnetic Fermi-liquid state
is equal to the sum of $s$ 
and $f$ electrons \cite{Mar82}: 
\begin{align}
N_{s}+ N_{f} =
2  \sum_{q k} \theta(\mu-\eta_{k}^q)\equiv V_{\rm F} \;.
\end{align}
$\eta_{k}^q$ ($q=1,2$) are the eigenvalues of the matrix
\begin{align}
\left( \begin{array}{cc} \epsilon_k & V_k \\[1ex] V_k^\ast & \epsilon_f 
    + \Sigma_k(0) \end{array} \right) \;,
\end{align}
where $\Sigma_k(\omega)$ is the proper selfenergy of the PAM.
Straightforward rearrangements 
lead to: 
\begin{align}
&N_{s}+ N_{f} 
= 2 \sum_k \left[ \theta(\mu-\epsilon_k) + 
\theta\left(\alpha-\frac{|V_k|^2}{(\mu - \epsilon_k)}\right)\right]
\label{eq:Ns_plus_Nf_allg}
\\[1ex]
&\mbox{with} 
\hspace*{2em}
\displaystyle \alpha={\mu-\epsilon_f-\Sigma_k(0)} \;.
\label{eq:alpha}
\end{align}
Eqs.~(\ref{eq:Ns_plus_Nf_allg}) and (\ref{eq:alpha}) are 
analogous to the ones obtained 
for the special case of infinite dimensions \cite{MaO95}.
As in Ref.~\onlinecite{MaO95} 
the three cases of less, equal, and more than half filling have to be
distinguished:  
\begin{align}
& (\alpha<0)
\hspace*{1.5em}
N_s + N_f = 2 \sum_k \theta(\mu-\epsilon_k - \Sigma_{s,k}^{}(0))
\label{eq:Lutt_lesshalf}
\\
& (\alpha=0)
\hspace*{1.5em}
N_s + N_f = 2 N
\label{eq:Lutt_equalhalf}
\\
& (\alpha>0)
\hspace*{1.5em}
N_s + N_f = 2 N + 2 \sum_k \theta(\mu-\epsilon_k - \Sigma_{s,k}^{}(0)) 
\label{eq:Lutt_morehalf}
\end{align}
$\Sigma_{s,k}^{}$ is the $s$-electron selfenergy as defined by 
an appropriate Dyson equation of the $s$-electron Green's function,
\begin{align}
G_{s,k}^{}(\omega) = \frac{1}{\omega - \epsilon_k + \mu -
  \Sigma_{s,k}^{}(\omega)}
\;.
\label{eq:G_ck}
\end{align}
It is 
related to the proper selfenergy $\Sigma_k(\omega)$ by
\begin{align}
\Sigma_{s,k}^{}(\omega) = \frac{|V_k|^2}{\omega -\epsilon_f +\mu -
  \Sigma_k(\omega)}\;.
\end{align}
According to the exact mapping of the PAM to the Kondo lattice model 
the $s$-electron selfenergy of the PAM becomes identical to
an analogously defined $s$-electron selfenergy of
the KLM in the extended Kondo limit 
\begin{align}
\Sigma_{s,k}^{} \;\stackrel{\mathrm{EKL}}{\longrightarrow}\;
\Sigma_{s,k}^{\mathrm{KLM}} \;.
\end{align}
Hence,
Eqs.~(\ref{eq:Lutt_lesshalf})-(\ref{eq:Lutt_morehalf}) with
$\Sigma_{s,k}^{}$ replaced by $\Sigma_{s,k}^{\mathrm{KLM}}$ represent
the analogue of the Fermi-surface sum rule for a non-magnetic
Fermi-liquid state of the Kondo lattice model.
The Fermi volume includes the number $N_f$ of 
localized spins. 
It is clear that for a magnetic Fermi-liquid state the corresponding
Fermi-surface sum rule of the PAM 
similarly maps to the Kondo lattice model.

In summary,
we have proven the exact mapping of the periodic Anderson model in the
extended Kondo limit to the Kondo lattice model 
for any coupling constant $J$.
As a consequence a direct proof of the large Fermi volume
of the KLM for a nonmagnetic Fermi-liquid state could be given.
Based on the mapping, further analytical and computational results 
for the KLM can be obtained in future.

This work was supported by Deutsche Forschungsgemeinschaft (Sfb 290). We
would like to thank M.~Potthoff for valuable discussions and comments.

\end{document}